\documentclass[aps,pre,reprint,floatfix,superscriptaddress]{revtex4-2}
\usepackage{graphicx}
\usepackage[latin1]{inputenc}
\usepackage{amsmath}
\usepackage{amsfonts}
\usepackage{amssymb}
\usepackage{xcolor}

\begin{document}

\title{External field-induced dynamics of a charged particle on a closed helix}

\date{\today}

\author{Ansgar~Siemens}
\email{asiemens@physnet.uni-hamburg.de}
\affiliation{Zentrum f\"ur Optische Quantentechnologien, Fachbereich Physik, Universit\"at Hamburg, Luruper Chaussee 149, 22761 Hamburg Germany} 
\author{Peter~Schmelcher}
\email{pschmelc@physnet.uni-hamburg.de}
\affiliation{Zentrum f\"ur Optische Quantentechnologien, Fachbereich Physik, Universit\"at Hamburg, Luruper Chaussee 149, 22761 Hamburg Germany}
\affiliation{Hamburg Center for Ultrafast Imaging, Universit\"at Hamburg, Luruper Chaussee 149, 22761 Hamburg Germany}

\begin{abstract}

\noindent We investigate the dynamics of a charged particle confined to move on a toroidal helix while being driven by an external time-dependent electric field. 
The underlying phase space is analyzed for linearly and circularly polarized fields. 
For small driving amplitudes and a linearly polarized field, we find a split-up of the chaotic part of the phase space which prevents the particle from inverting its direction of motion.
This allows for a non-zero average velocity of chaotic trajectories without breaking the well-known symmetries commonly responsible for directed transport. 
Within our chosen normalized units, the resulting average transport velocity is constant and does not change significantly with the driving amplitude. 
A very similar effect is found in case of the circularly polarized field and low driving amplitudes. 
Furthermore, when driving with a circularly polarized field, we unravel a second mechanism of the split-up of the chaotic phase space region for very large driving amplitudes. 
There exists a wide range of parameter values for which trajectories may travel between the two chaotic regions by crossing a permeable cantorus. 
The limitations of these phenomena, as well as their implication on manipulating directed transport in helical geometries are discussed.

\end{abstract}

\maketitle

\section{Introduction}

Helical structures and patterns can be frequently found in nature, with systems ranging from molecules like DNA or amino acids to self-assembled configurations of particles in nanotubes \cite{vernizzi2009}. 
Especially for quasi 1D structures, the helical geometry can offer advantages such as increased stability and resistance to deformations \cite{marko1994a, kornyshev2007a}. 
In the last decades great progress was made in attempts to synthesize artificial 1D nanostructures, such as helical CNT's \cite{lau2006a}, with hopes for applications in nano electronic circuits \cite{avouris2006a, avouris2008a, hatton2008a, shaikjee2012a}. 
Therefore, there is a great interest in understanding how the electronic properties of 1D structures are affected by helical geometries.

Already in minimal models, intriguing phenomena can result from the geometric confinement to a 1D helix. 
It was demonstrated that, due to the geometry, ballistic long-range Coulomb interacting particles on a 1D helical path can form bound states \cite{kibis1992a,schmelcher2011} and can even build 1D lattice structures \cite{schmelcher2011,siemens2020,zampetaki2018}.
Novel physics resulting from this behavior has been reported in several works discussing relevant setups \cite{zampetaki2013, zampetaki2015, zampetaki2015a, pedersen2014a, pedersen2016c, pedersen2016d, zampetaki2017, plettenberg2017a}. 
Effects range from mechanical properties like an unusual electrostatic bending response \cite{zampetaki2018}, to intriguing nonlinear dynamics, such as the scattering of bound states at an inhomogeneity in the 1D path \cite{zampetaki2013} or the tuning of the dispersion relation of a 1D chain of bound particles by varying the helix radius \cite{zampetaki2015}. 
In the latter example, a degeneracy of the band structure for a specific helix radius was identified which prevents excitations from dispersing through the system.

In helical systems, the novel effects typically emerge due to the fact that the acting forces are partially compensated by confining forces of the helix, and are therefore not limited to Coulomb interactions. 
Effects of dipole-dipole interactions \cite{pedersen2014a, pedersen2016c, pedersen2016d}, as well as external electric fields \cite{plettenberg2017a,siemens2020} have been explored. 
Previous investigations of external electric fields considered adiabatically varying forces and demonstrated the possibility of using an external electric field for controlled state transfer \cite{plettenberg2017a}, and inducing crystalline lattice ordering of particles \cite{siemens2020}.
For crystalline particles on a closed helix exposed to a static electric field, an unconventional pinned-to-sliding transition has been observed \cite{zampetaki2017}. 
Investigating the dynamics of confined particles being driven by time-dependent external forces is therefore a natural next step.

Periodic driving is at the core of many intriguing phenomena, such as resonances and chaos. 
In driven systems, already simple models can often yield quite complex dynamics and give valuable insight into real physical systems. 
For example, the model of a driven morse-oscillator can give insight into the (vibrational) stability of molecules \cite{goggin1988}. 
In the same spirit, particles in driven double well potentials have been studied to explain the tunneling dynamics (or the suppression thereof) through a potential barrier \cite{buttiker1982, pimpale1991, bagwell1992, ge1996, mateos1998, mateos1999, leonel2004}. 
Studies of driven Hamiltonian systems, i.e. particle dynamics in time dependent periodic potentials, often possesses a focus on the manipulation of transport phenomena due to the choice of the driving potential. 
There, the transport of diffusive trajectories is usually induced by breaking certain spatio-temporal symmetries \cite{flach2000, schanz2001, quintero2010, reimann2002a, hanggi2009, denisov2014, renzoni2009}.
However, other manipulation techniques, like the possibility of switching between ballistic and diffusive motion by introducing localized disorder \cite{wulf2014}, have been demonstrated. 
Furthermore, the presence of spatially varying forces has been linked to a variety of intriguing phenomena, such as the formation of density waves \cite{petri2011}. 
Based on this understanding of driven systems, a plethora of applications, including velocity filters \cite{petri2011, wulf2012}, spectrometers \cite{wambaugh2002, mukhopadhyay2018, matthias2003}, or batteries extracting energy from thermal fluctuations \cite{reimann2002, hanggi2009, denisov2014, matthias2003, schmitt2015, serreli2007, cubero2016, astumian1994, astumian1997}, have been proposed.

Motivated by the complexity arising when particles are confined to curved space, we investigate in this work the influence of time periodic forces on particles in helical confinement. 
As a prototype, we consider a single particle confined to a toroidal helix, being driven by either an oscillating or a rotating electric field.
The combination of driving and confining forces leads to spatially and temporally varying effective forces. 
For a wide range of driving amplitudes, the systems phase-space resembles that of a particle in either a standing wave (oscillating driving field) or a running wave (rotating driving field). 
However, for very low driving amplitudes, as well as for large driving amplitudes in case of the oscillating field, we identify two different scenarios by which the chaotic phase space region can be split. 
We explain how these splits are induced by the different scales of oscillations in the driving potential, and how they influence the corresponding transport phenomena.

Our manuscript is structured as follows.
Sec. \ref{sec:theory} contains the parametrization of the toroidal helix, a discussion of the Lagrangian, and the general equations of motion for our setup. 
We further discuss the considered driving laws.
In Sec. \ref{sec:driveX} and \ref{sec:driveXY} we investigate and analyze the dynamics int the presence of driving with a linearly polarized and a circularly polarized electric field respectively. 
Finally, in Sec. \ref{sec:end} we provide our conclusions.

\section{Particles in Helical Geometries with External Driving}
\label{sec:theory}

\begin{figure}
\includegraphics[width=\columnwidth]{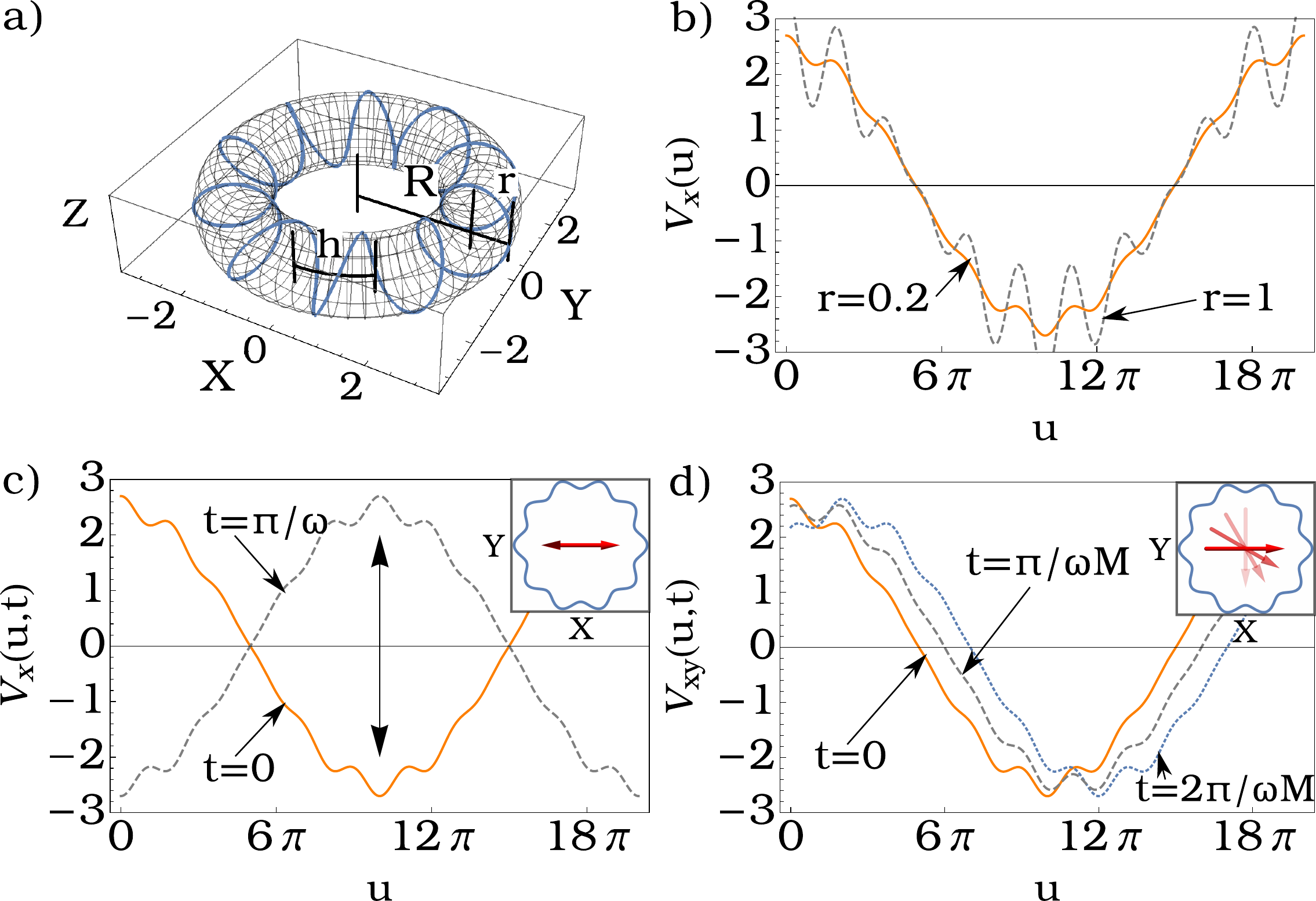}
\caption{\label{figure1} (a) A 3D illustration of the parametric function $\textbf{r}(u)$, for $M=10$, $r=0.8$, and $R=2.5$. (b) The potential $V_x(u)$ created by a static field in the x-direction shown for toroidal helices with $M=10$, $R=2.5$, and helix radii of $r=0.2$ (orange) and $r=1$ (dotted gray).(c) The potential landscape $V_x(u,t)$ for a linearly polarized oscillating field in x-direction shown for $t=0$ (orange) and $t=\pi/\omega$ (dashed gray). The inset in the top right corner visualizes the driving direction of the field (red) for a top view of the setup. (d) The potential landscape $V_{xy}(u,t)$ for driving with a circularly polarized field in the xy-plane shown for the times $t=[0,\pi/\omega M, 2\pi/\omega M]$. After the time $t=2\pi/\omega M$ the motion of the potential repeats, being shifted by $\Delta u = 2\pi$. Again, the inset in the top right corner visualizes the driving field (red) for a top view of the setup.}
\end{figure}

We consider a single particle with charge $q$ confined to move along a toroidal helix (see Fig. \ref{figure1}(a) for a visualization). 
The parametrization of the particles positions is then given by the following equation
\begin{equation}\label{eq:1}
\textbf{r}(u):=\left(
\begin{array}{c}
\left(R+r\cos(u)\right)\cos(u/M) \\
\left(R+r\cos(u)\right)\sin(u/M) \\
r\sin(u)
\end{array}
\right),
u\in [0,2\pi M]
\end{equation}
where $R$ is the torus radius determining how strongly our helix is bent, $r$ is the radius of the helix, and $M$ are the total number of helical windings.
Since the path is closed we have $\textbf{r}(u) = \textbf{r}(u+2\pi M)$, and the parameters obey the following restriction $R=M h/2\pi$, where $h$ is the pitch of the helix.
When $u$ changes by an amount of $2\pi$, the particle moves the distance of one winding on the helix.
When $u$ changes by an amount of $2\pi M$, the particle circles once around the torus and is exactly at the same position it started in.

The driving force is assumed to be caused by an external electric field $\textbf{E}$. 
The potential energy $V(u,t)$ of the particle is then given by
\begin{equation}
V(u,t)=q\textbf{E}(t)\cdot\textbf{r}(u)
\end{equation}
Our system is then described by the following Lagrangian
\begin{equation}\label{eq:lagrangian1}
\mathcal{L}=\dfrac{m}{2}\left(\dfrac{d\textbf{r}(u)}{dt}\right)^2-q\textbf{E}\cdot\textbf{r}(u)
\end{equation}
Equation (\ref{eq:lagrangian1}) already accounts for the confining forces of the setup by only allowing positions $\textbf{r}(u)$ on the parametric helical curve.
Since $\textbf{r}(u)$ is known, we can already evaluate the derivative in the kinetic energy term and rewrite Eq. \ref{eq:lagrangian1} as
\begin{equation}
\mathcal{L}=\dfrac{m}{2}\zeta(u)\left(\dfrac{du}{dt}\right)^2 -q\textbf{E}\cdot\textbf{r}(u)
\end{equation}
where $\zeta(u):=(d\textbf{r}(u)/du)^2=r^2+\left(R+r\cos(u)\right)^2/M^2$.
From this, we obtain the following equations of motion for an arbitrary driving field $\textbf{E}(t)$
\begin{equation}\label{eq:eom2}
\zeta(u)\dfrac{d^2u}{dt^2} =  \dfrac{1}{2} \dfrac{d\zeta(u)}{du} \left(\dfrac{du}{dt}\right)^2 - \dfrac{q\textbf{E}(t)}{m}\dfrac{d\textbf{r}(u)}{du}
\end{equation}
%
%

Let us now take a closer look at the driving potential $V(u,t)=q\textbf{E}(t)\cdot\textbf{r}(u)$ created by the electric field. 
Even in the static case, i.e. without time dependence, the potential can become quite complex and possesses multiple minima. 
This is shown in Fig. \ref{figure1}(b) for a static field parallel to the x-axis.
In this case the electric field is given by $\textbf{E}=E\textbf{e}_x$ and the potential energy becomes
\begin{equation}
V_x(u) = qE \left(R+r \cos(u)\right)\cos(u/M)
\label{Ex_static}
\end{equation}
This potential consists of two terms: 
The $R\cos(u/M)$ term creates a long wavelength cosine shaped potential that is maximal at the position that extends most into the x-direction (for $u=0$ or $u=2\pi M$) and minimal for the position extending most into the negative x-direction (for $u=\pi M$). 
Since it is caused by the overall toroidal shape of the curve $\textbf{r}(u)$ we will call this the \textit{torus induced potential} (TIP).
On top of that, there is a smaller modulation given by the $r\cos(u)\cos(u/M)$ term. 
Since this modulation originates from the helix windings we will call this the \textit{winding induced potential} (WIP).
The amplitude of the WIP can be modulated via the helix radius $r$ (shown in Fig. \ref{figure1}(b) for $r=0.2$ and $r=1$). 
Due to the $\cos(u/M)$ dependence, the WIP oscillational amplitude also changes with the position on the torus.
The amplitude is largest for $u\in[0,\pi M,2\pi M]$ and vanishes for $u\in[\pi M/2,3\pi M/2]$.
The number of minima in the modulation is determined by the number of helical windings $M$.

In this work we focus on two different time dependent fields:
Driving with a field oscillating parallel to the x-axis, and driving with a field rotating in the xy-plane. 
In the first case, the driving field becomes $\textbf{E}(t)=E\cos(\omega t)\textbf{e}_x$. 
The resulting potential $V_x(u,t)$ is a standing wave with the shape shown in Fig. \ref{figure1}(c)
\begin{equation}\label{eq:drivingLaw}
V_x(u,t) = qE \left(R+r \cos(u)\right)\cos(u/M)\cos(\omega t)
\end{equation}

When we consider an electric field rotating in the xy-plane the driving becomes slightly more complex. 
In this case, the electric field can be written as $\textbf{E}(t)=E\{\cos(\omega t),\sin(\omega t),0\}$ and the potential landscape becomes
\begin{equation}\label{eq:drivingLaw2}
V_{xy}(u,t) = -qE\cos(\omega t - u/M)\left(R+ r \cos(u)\right)
\end{equation}
Figure \ref{figure1}(d) visualizes the time evolution of this potential by showing the potential landscape at different times $t$. 
The three curves in the figure correspond to the cases $t=0$ (orange), $t=\pi/\omega M$ (gray), and $t=2\pi/\omega M$ (blue). 
Due to the symmetries of the toroidal helix, we only need to consider the time $\Delta t=2\pi/\omega M$ needed to rotate by one winding to understand the driving, since the potential movement repeats after this time; it is just shifted by a distance of $\Delta u=2\pi$.
The time evolution of the potential landscape resembles a `crawling' motion: 
The local extrema of the potential oscillate between being a potential minimum and a potential maximum, with a constant phase shift of $2\pi/M$ between neighboring minima (or maxima). 
A video showing the time evolution of $V_{xy}(u,t)$ can be found in the supplementary material \footnote{See Supplemental Material at [URL will be inserted by publisher] for a video of the time evolution of Eq. \ref{eq:drivingLaw2}.}. 
It should also be noted, that for $V_{xy}(u,t)$ the equations of motion are not symmetric with regards to the spatio-temporal symmetries given by ($u\rightarrow-u+\Delta u$, $t\rightarrow t+\tau$) and ($u\rightarrow u+\Delta u$, $t\rightarrow-t+\tau$); a necessary criterion for directed transport within the chaotic sea \cite{flach2000, quintero2010}. 
In contrast, these symmetries are conserved for $V_x(u,t)$.

We can eliminate redundant parameters by introducing dimensionless units.
Without loss of generality, we choose to express distances in units of $2h/\pi$ and time in units of $\omega/2\pi$. 
We also normalize the particle mass and charge to $m=q=1$ (which is the same as `absorbing' both values in the driving amplitude).
The remaining independent system parameters are the winding number $M$, the helix radius $r$, and the driving amplitude $E$. 

A final remark on our computational approach is in order: 
The equations of motion are numerically integrated with the Dormand-Prince method - a Runge-Kutta method with variable step size. 
The maximal step-size of our time steps was chosen as $\Delta t=0.01$. 
It was verified that this maximum step-size produces accurate results even for driving amplitudes as large as $E>2000$, which is much larger than any driving amplitude used in this work.

\section{Particle Dynamics for a Linearly Polarized Field}
\label{sec:driveX}

\begin{figure*}
\includegraphics[width=\linewidth]{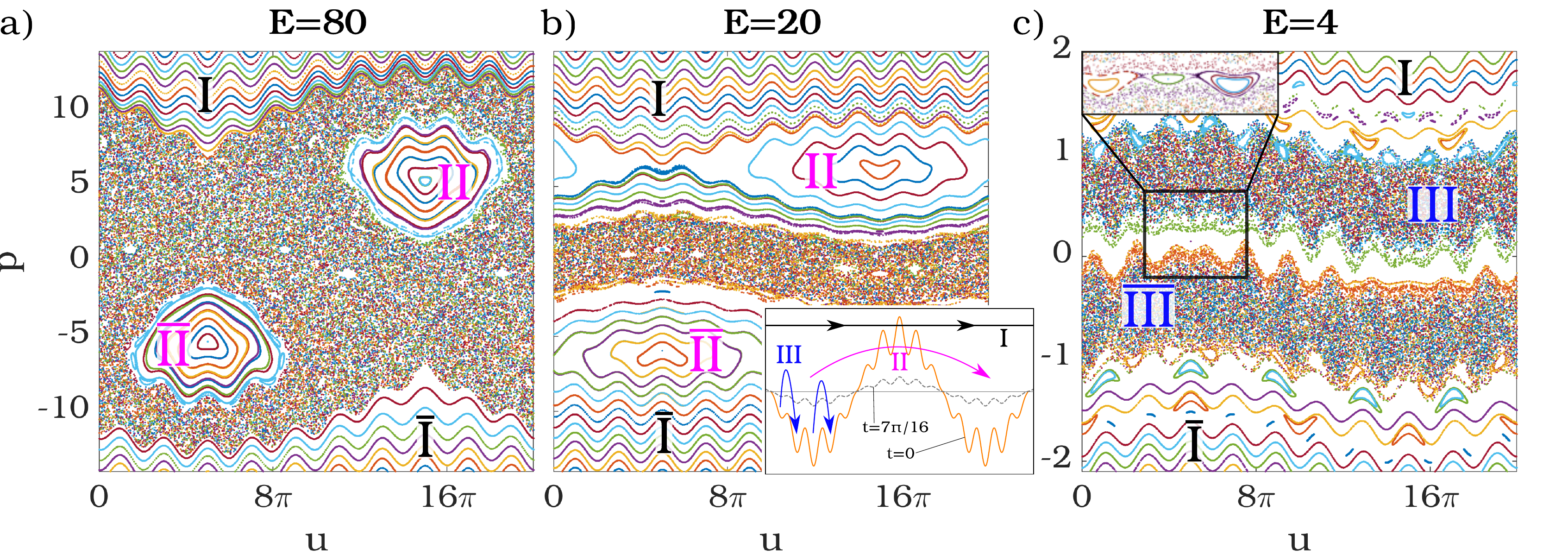}
\caption{\label{figure2}  Poincar\'e surfaces of sections (PSOS) for a particle on the toroidal helix, driven by a linearly polarized oscillating field for (a) $E=80$, (b) $E=20$, and (c) $E=4$. 
Different colors are assigned to the trajectories for easier differentiation.
Each PSOS features between $45$ and $75$ trajectories, each simulated for $2000$ driving periods. 
(Quasi-) periodic trajectories between the two chaotic regions (around $p=0$) in (c) are only shown in the inset (top left of (c)) to emphasize the splitting of the chaotic sea into two parts in the main figure. 
The inset in (b) visualizes the particle motion on $V_x(u)$ for the three different types of trajectories (I-III) during a driving period in the range $u\in[-2\pi M/3, 2\pi M/3]$.  
The symbols $\overline{\text{I}}$-$\overline{\text{III}}$ mark the inverse of trajectories (I-III), i.e. trajectories moving in the opposite direction. 
(I) `Quasi-free' trajectories that are too fast to be significantly affected by the driving. (II) Trajectories belonging to the large regular islands with the fixed points at $|p|\approx5.35$ which move around the torus once during every driving period. (III) The chaotic trajectories (after the chaotic sea has split) move by one winding during each driving period. }
\end{figure*}

In this section we will analyze the dynamics when the system is driven by an electric field oscillating parallel to the x-axis. 
For this we will examine the phase space of the system and understand how it is decomposed for different parameter regimes. 
The dimensions of the phase space are made up of the three parameters: position $u$, momentum $p$, and time $t$.
Since our Lagrangian is periodic in time, we can use a Poincar\'e surface of sections (PSOS) - specifically a stroboscopic map - to visualize the phase space in a two dimensional stroboscopic $u(p)$ dependence. 
Note, that our momentum $p$ refers to the canonical momentum given by
\begin{equation}
p=\dfrac{du/dt}{m\left( r^2 + ((R+r\cos(u))/M)^2 \right)}
\end{equation}

We start our investigation by considering a toroidal helix with $M=10$ and $r=0.2$. 
Figure \ref{figure2} shows the PSOS of the system for electric field amplitudes $E=80$, $E=20$, and $E=4$. 
As we will see, the phase space for large and intermediate driving amplitudes will closely resemble that of a particle in a standing wave \cite{menyuk1985}. 
However, for low driving amplitudes, we observe novel features of the dynamics arising from the interplay of WIP and TIP. 
The investigation of these dynamics and their implication for manipulating directed transport will be the main result of this section.  

In Fig. \ref{figure2}(a), for $E=80$, we observe a mixed phase space that mainly allows three different kinds of trajectories: 
Chaotic trajectories, and two types of (quasi-) periodic trajectories which we will refer to as \textit{Type-I} and \textit{Type-II} trajectories. 
Type-I trajectories (marked I and $\overline{\text{I}}$ in the figure) are invariant spanning curves \cite{saito1982, loskutov2000} for which the particle momentum is too large to be significantly affected by the driving. 
The driving results only in a weak modulation of their dynamics. 
Towards smaller momenta the Type-I trajectories border on a `sea' of chaotic trajectories which contains two large regular islands. 
These regular islands correspond to the Type-II trajectories and describe motion around the torus in phase with the driving period, i.e. after one driving period the particle on a Type-II trajectory has circled the torus exactly once. 
Both regular islands describe the same kind of motion, but in opposite directions.

As one might expect, the size of the chaotic portion of phase space decreases when the driving amplitude is decreased. 
This can be seen in Fig. \ref{figure2}(b) where $E=20$. 
The Type-I trajectories, as well as the two main fixed points we identified in the previous figure are still present. 
However, the chaotic region now occupies a much smaller momentum range of the phase space. 
In addition, at the center of the chaotic region around $p\approx0$ additional fixed points appear in the phase space, e.g. at $[u,p]\approx[10.5,0.45],[15.7,0.8]$ and $[20.7,0.45]$. 
They correspond to initial conditions in which the particle stays within a narrow range of $u$ and is hardly affected by the driving. 
The reason for their appearance is as follows: 
When the driving amplitude decreases, so does the acceleration of the particle. 
Below a certain threshold the particle has hardly moved before the driving field accelerates the particle in the opposite direction. 
With decreasing driving amplitude an increasing amount of trajectories with initial conditions around $p=0$ will exhibit this behavior. 
The effect on the phase space can be seen in Fig. \ref{figure2}(c) for $E=4$ (note the adjusted range of $p$ values). 
Here the driving amplitude is sufficiently small, such that for every $u$ there is a (quasi-) periodic trajectory (pictured only in the inset of (c)) close to $p=0$ that is hardly affected by the driving and mostly stays in place. 
An interesting result is, that the appearance of these trajectories is splitting the chaotic sea into two parts: 
One with $p>0$ (marked III in Fig. \ref{figure2}(c)) and one with $p<0$ (marked $\overline{\text{III}}$) which we will refer to as Type-III trajectories. 
This has significant consequences for the dynamics. 
Type-III trajectories starting in the chaotic region with $p>0$ will remain there and maintain a strictly positive momentum. 
Inverting the direction of movement is impossible, since that requires slowing down and crossing the region of regular islands around $p=0$. 
The same is of course true for trajectories starting in the chaotic region with $p<0$. 
In other words: 
%
%
When the chaotic sea splits up, we transition from a single chaotic sea in which all trajectories have an average velocity of zero, to two completely separated (symmetric) chaotic seas in which chaotic trajectories have an average velocity of either $+2\pi$ (upper chaotic sea) or $-2\pi$ (lower chaotic sea).

\begin{figure}
\includegraphics[width=\columnwidth]{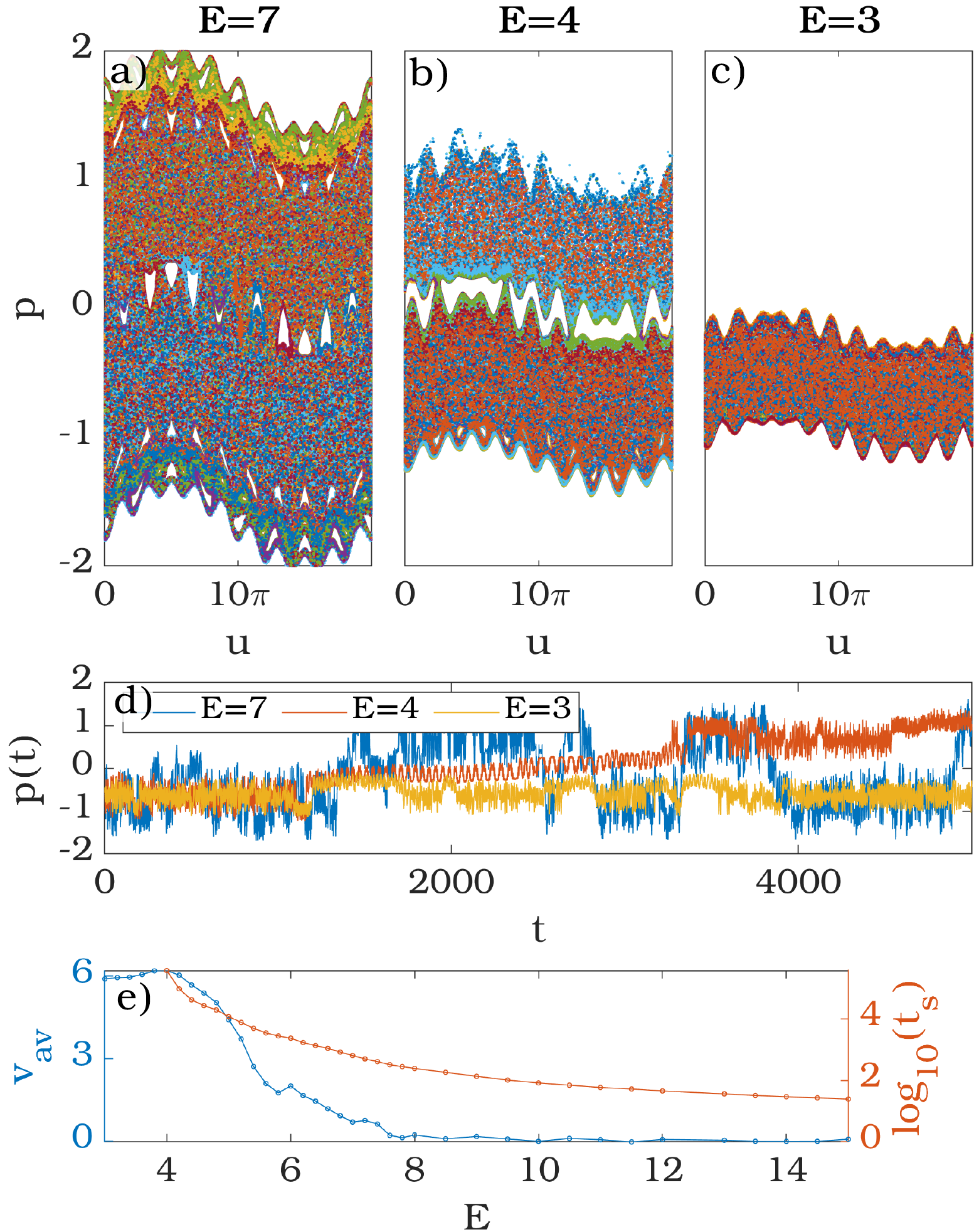}
\caption{\label{figure3} (a-c) PSOS created from $\sim10^3$ trajectories with initial conditions in the chaotic sea ($u\in[0, 1]$ and $p\in[-0.42, -0.62]$). Each particle was simulated for $5000$ driving periods. The appearance of stable trajectories around $p=0$ splits the chaotic sea into two seas, when the driving amplitude $E$ is decreased. (d) Representative example trajectories emanating in the chaotic sea for $p<0$ for $E=3$ (yellow), $E=4$ (orange), and $E=7$ (blue). The two momentum regimes the particles are confined to are clearly visible. Transition between the two regimes is more likely for larger driving amplitudes. (e) The average transport velocity $v_{av}$ and switch time $t_s$ as a function of the driving amplitude. Each data point was obtained from simulation numbers and times similar to those of (a-c).  }
\end{figure}

In Fig. \ref{figure3} we take a closer look at this split-up of the chaotic sea.
A close up of the split appearing in the PSOS is shown in Fig. \ref{figure3}(a-c). 
For clarity, the PSOS's of Fig. \ref{figure3}(a-c) only contain initial conditions from the chaotic region with $p<0$. 
In Fig. \ref{figure3}(a), at $E=7$, the emerging (quasi-) periodic regions around $p=0$ are clearly visible. 
However, changing the direction of motion is still possible and happens indeed frequently. 
The momentum evolution $p(t)$ of a representative example trajectory is shown in Fig. \ref{figure3}(d)(blue curve). 
From this $p(t)$ curve we can see that already for $E=7$ there are effectively two momentum ranges the particle can be confined to. 
The particle frequently switches between having either positive or negative momentum for extended periods of time. 

When the driving amplitude is decreased further to $E=4$ (see Fig. \ref{figure3}(b)), the two chaotic phase space regions are almost separated from each other.
An inversion of the direction of movement now happens much less frequently. 
In the phase space this can be seen from the decreasing density for $p>0$. 
From the corresponding example trajectory in Fig. \ref{figure3}(d)(red curve), we see that the momentum inversion now also takes a much longer time than for $E=7$. 
It takes our example trajectory almost $\sim2200$ driving periods to change its momentum from $p<0$ to $p>0$.

Finally, for $E=3$ (Fig. \ref{figure3}(c)), the two phase space regions are completely separated from each other. 
None of our trajectories cross into the phase space region with $p>0$.  
In this regime, the dynamics of all simulated trajectories resemble that of our example trajectory in Fig. \ref{figure3}(d)(yellow curve): 
The trajectories are chaotic while sustaining a strictly negative momentum.

A better understanding of the Type-III trajectories can be gained from statistical averages. 
We consider the average velocity $v_{av}$, as well as the \textit{mean switch time} $t_s$. 
For a set of trajectories $u(u_i,p_i,t)$ with initial conditions $u(t=0)=u_i$ and $p(t=0)=p_i$ the average velocity is determined by averaging the mean velocities of all trajectories 
\begin{equation}
v_{av} = 
\dfrac{1}{N T} \sum_{i=1}^N \int_0^T \dfrac{du(u_i,p_i,t)}{dt} dt
\end{equation}
where $T$ is the simulation time of individual trajectories. 
We define the mean switch time as the average time a particle spends with $p>0$ (or $p<0$) before inverting the direction of its motion. 
Note, that within our numerical simulations, there are limitations regarding the calculation and accuracy of $t_s$. 
We can only determine $t_s$ accurately from our simulations, if we (on average) observe at least one switch in the time $T$. 
Since each trajectory was simulated for $T=5000$ driving periods, our value of $t_s$ is accurate for values below $t_s\lesssim2500$. 
In practice, we simulate $10^3$ trajectories for $5000$ time steps, count the total number of switches $n$ in all simulations, and then calculate $t_s=0.5\ 10^7/n$.

Figure \ref{figure3}(e) shows both $v_{av}$ and $t_s$ as a function of the driving amplitude. 
For better insight into the dynamics of the Type-III trajectories both curves were only obtained from trajectories with $p(t=0)>0$.
Until the split-up of the chaotic region at about $E=4$ both quantities increase with decreasing driving amplitude. 
From $t_s$ we see that long before the two chaotic regions are separated from each other, the particles perform very long 'flights' without inverting the direction of their motion. 
Even for $E=7$, where the chaotic regions are still reasonably well connected in the phase space (see Fig. \ref{figure3}(a)), we have a mean switch time of $t_s>600$ driving periods.

In the figure, our mean switch time exceeds the critical value of $t_s=2500$ for driving amplitudes $E<6$. 
As stated above, we cannot accurately calculate $t_s$ in this regime of driving amplitudes because the change of the direction of motion happens too infrequently. 
Consequently, in this regime the choice of initial conditions ($p(t=0)<0$) becomes apparent in the statistics of $v_{av}$. 
While $t_s>2500$, the average velocity $v_{av}$ increases with increasing $t_s$. 
When the two chaotic phase space regions splits up at around $E=4$, $v_{av}$ reaches a plateau (see $v_{av}$ in Fig. \ref{figure3}(e)). 
After the split-up, the Type-III trajectories have a consistent mean velocity of slightly less than $v_{av}\approx2\pi$. 
This velocity corresponds to a position change of about one helix winding during each driving period. 
More precisely, each driving period the chaotic Type-III trajectories move between neighboring minima in the WIP. 
Therefore, the dynamics of Type-III trajectories are similar to the Type-II trajectories, except that they are mostly determined by the minima of the WIP with the TIP being a perturbation that is mostly responsible for the chaos. 
In contrast, the Type-II trajectories are mostly determined by the minima of the TIP, with the WIP acting as a perturbation. 
For even lower driving amplitudes the perturbation due to the TIP becomes small enough for the Type-III trajectories to stabilize into a series of fixed points (similar to the ones shown in Fig. \ref{figure4}(a) for a rotating driving field).

Since the Type-III trajectories emerge due to the WIP, it is no surprise that the occurrence of the phase space split depends on the helix radius $r$. 
For larger values of $r$, the (quasi-) periodic trajectories around the Type-III fixed points will already stabilize for larger values of $E$, since the relative strength of the `perturbation' due to the TIP decreases. 
For a large enough $r$, it is possible for the Type-III fixed points to stabilize before the chaotic region is splitting up. 
In extreme cases this may even prevent the occurrence of chaotic Type-III trajectories.

The only independent system parameter did not discuss so far is the winding number $M$. 
Changing $M$ does not significantly affect the overall dynamics. 
However, due to the relation $R=Mh/2\pi$ and our choice of units (thereby setting $h=\pi/2$), changing $M$ will change the torus radius $R$, thereby changing the momentum of the Type-II trajectories. 
This, in turn, changes e.g. the driving amplitude required for a mixed phase space as shown in Fig. \ref{figure2}(a). 
This also changes the ratio of $r/R$ and may cause the periodic Type-III fixed points to stabilize at different driving amplitudes. 
Increasing $M$ also increases the number of extrema in the WIP, leading to more fixed points in the (quasi-) periodic Type-III trajectories once they stabilize. 
Besides this, however, the split-up of the chaotic phase space region is mostly unaffected.

\section{Particle dynamics in the presence of a circularly polarized field}
\label{sec:driveXY}

\begin{figure}
\includegraphics[width=\columnwidth]{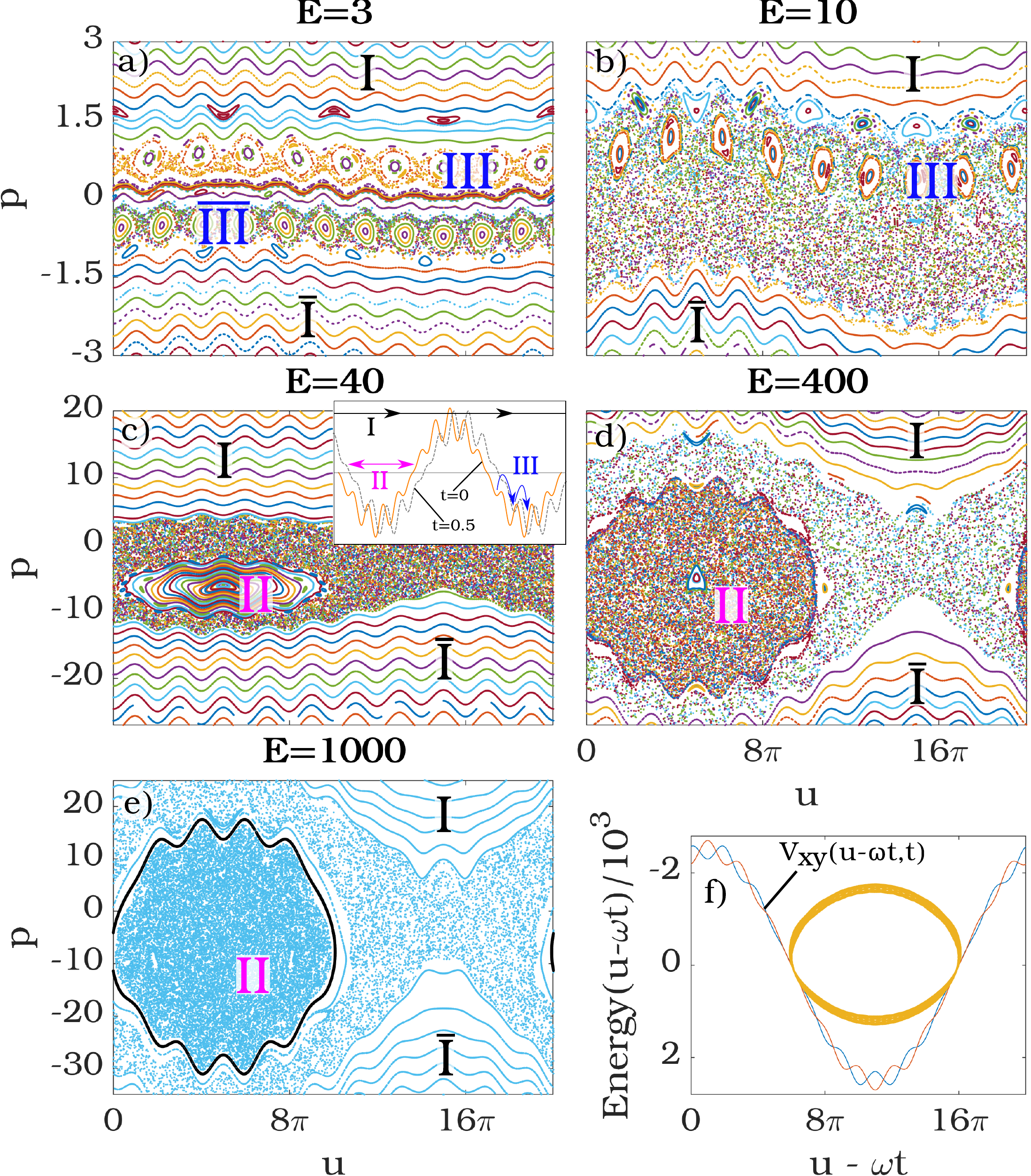}
\caption{\label{figure4} Poincare Surfaces of Sections (PSOS) for a particle on a toroidal helix driven by a circularly polarized field in the xy-plane. The inset in (c) visualizes the dynamics of different trajectories (I, II, and III) on $V_{xy}(u,t)$ in the range $u\in[-2\pi M/3, 2\pi M/3]$. 
The symbols $\overline{\text{I}}$-$\overline{\text{III}}$ mark the inverse of trajectories (I-III), i.e. trajectories moving in the opposite direction. 
(I) `Quasi-free' trajectories that are too fast to be significantly affected by the driving. (II) Trajectories that are trapped in a well of the potential $V_{xy}$ and move around the torus once during every driving period. (III) Chaotic and regular trajectories that (after the chaotic sea has split) move one helical winding during each driving period. 
The PSOS are shown for driving amplitudes of (a) $E=3$, (b) $E=10$, (c) $E=40$, (d) $E=400$, and (e) $E=1000$. In (e) the coloring was changed to emphasize the split of the chaotic region; except for one highlighted periodic trajectory (black) all data points are colored blue. The two chaotic regions correspond to chaotic motion that is trapped in the moving potential, and chaotic motion that is (on average) slower than the moving potential. The yellow curve in (f) shows the motion of the highlighted (black) trajectory of (e) in the moving potential.  }
\end{figure}

Another intriguing split-up in the phase space can be observed when driving with a circularly polarized field in the xy-plane. 
In this case, the driving law is characterized by the time dependent potential landscape $V_{xy}(u,t)$ given in Eq. (\ref{eq:drivingLaw2}). 
In this section, we will encounter trajectories that are very similar to the Type-I-III trajectories that were classified in Sec. \ref{sec:driveX}. 
We will again refer to them as Type-I, II, and III trajectories. 
Type-I trajectories are again invariant spanning curves that limit the momentum of chaotic trajectories and are hardly affected by the driving. 
Type-II trajectories move around the torus in phase with the driving. 
This time, however, the potential $V_{xy}$ describes a running wave, and the Type-II trajectories correspond to particles that are trapped in one of the moving potential wells. 
Type-III trajectories refer to trajectories that are unable to invert their direction of movement and move between successive minima of $V_{xy}$ during each driving period with an average velocity of $v_{av}=2\pi$.

An overview of the phase space for $M=10$ and $r=0.2$ is given in Fig. \ref{figure4}(a-e). 
For a large part, the phase space is very similar to the one shown in the previous section: 
There is a large regular island of Type-II trajectories corresponding to motion around the torus in phase with the driving period. 
The size of the corresponding chaotic region increases with the driving amplitude and leads to a mixed phase space for large $E$. 
The chaotic region is surrounded by Type-I trajectories.
Also, the $r \sin(u)$ dependence of $V_{av}$ leads to the presence of Type-III trajectories for very low driving amplitudes which, due to perturbations in form of the $R$ dependent term in $V_{av}$, can be chaotic and lead to a splitting of the chaotic sea similar to the one discussed in Sec. \ref{sec:driveX}. 
At the same time, however, there are major differences. 
Since our driving law breaks parity and time inversion symmetries in the equations of motion, the resulting phase space is not symmetric anymore. 
Instead of two fixed points with Type-II trajectories like in Fig. \ref{figure2}, there is now only one that corresponds to motion around the torus with the same direction as the rotation of the driving field . 
Furthermore, the emergence of Type-III trajectories with decreasing driving amplitude is not symmetric anymore. 
For our example parameters (quasi-) periodic Type-III trajectories with $p>0$ emerge even before the split of the chaotic sea has begun (see Fig. \ref{figure4}(b)).

The most interesting difference, however, emerges for very large driving amplitudes. 
Whereas in the case of a linearly oscillating driving field a larger driving amplitude mostly leads to an increased chaotic region, new structures can emerge in the phase space when driving with a rotating large amplitude field. 
For very large driving amplitudes (see Fig. \ref{figure4}(e)) regular (quasi-) periodic trajectories appear and split-up the chaotic sea into two regions. 
These (quasi-) periodic trajectories correspond to Type-II trajectories that move around the toroidal helix in phase with the driving. 
This can be seen from Fig. \ref{figure4}(f) which shows the path of the highlighted (black) trajectory from Fig. \ref{figure4}(e) in the driving potential. 
For convenience, the data is plotted in a moving reference frame that is moving in phase with the driving potential.

The two different chaotic regions in Fig. \ref{figure4}(e) correspond to different kinds of chaotic motion. 
The chaotic region surrounded by the newly stabilized periodic Type-II trajectories consists entirely of trajectories that are trapped in a valley of our driving potential. 
While the motion is chaotic, each trajectory will on average move in phase with the driving, once around the toroidal helix during each driving period. 
These trajectories are consequently also Type-II trajectories - just chaotic and not (quasi-) periodic. 
With increasing driving amplitude, the chaotic Type-II trajectories will stabilize into periodic Type-II trajectories.

\begin{figure}
\includegraphics[width=\columnwidth]{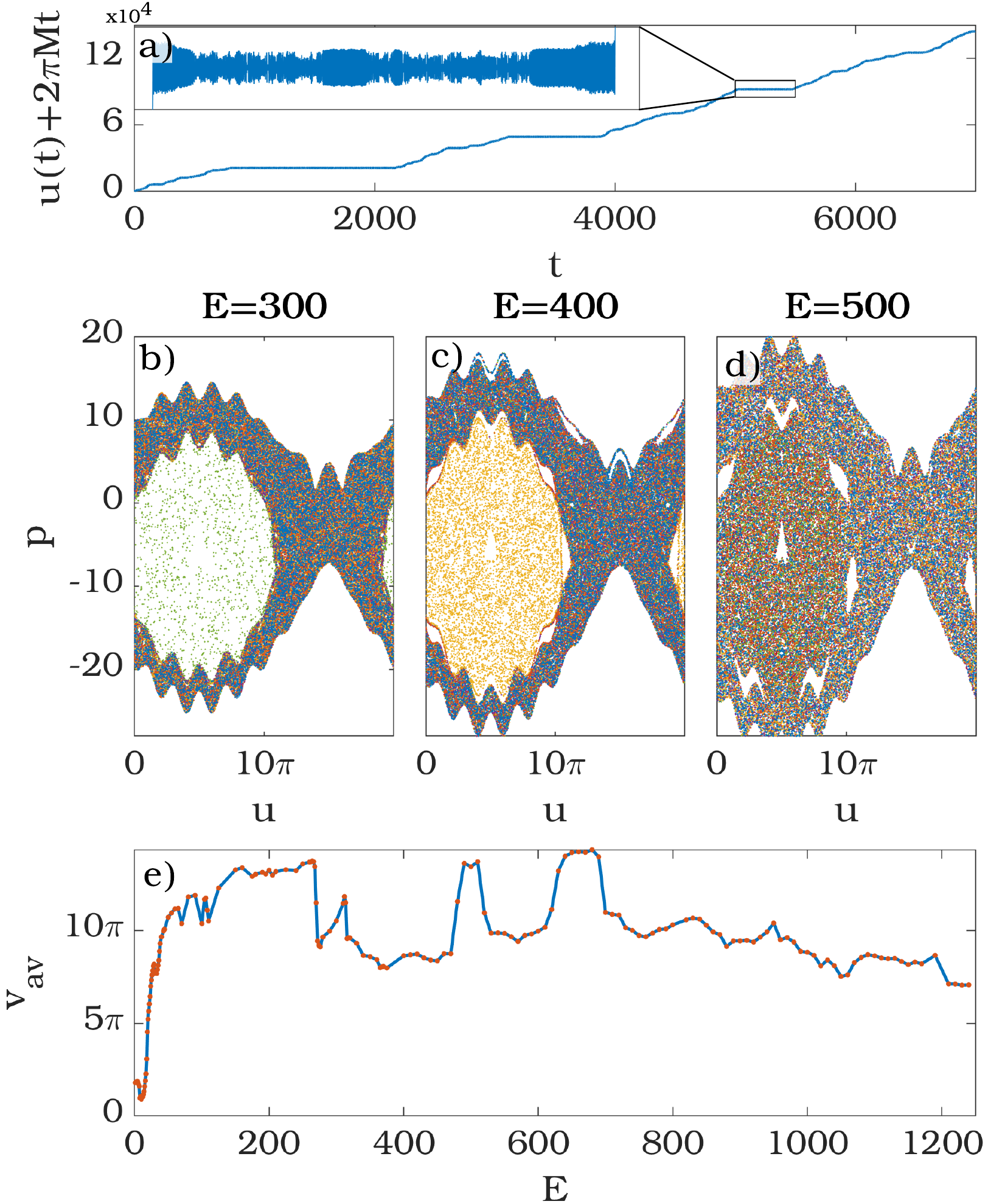}
\caption{\label{figure5} (a) Example trajectory in a comoving reference frame that moves in phase with the driving field. When the particle crosses the cantorus and becomes a chaotic Type-II trajectory, $u(t)-2\pi Mt$ will become constant, which is demonstrated in the inset. (b-d) Each figure shows a PSOS for six trajectories (with the same initial conditions for each figure) for (b) $E=300$, (c) $E=400$, and (d) $E=500$.  In (b) and (c) only one of the six trajectories manages to cross the cantorus, whereas in (c) all trajectories frequently switch between the two types of chaotic motion. (e) Average velocity for particles started in the chaotic sea with initial conditions chosen close to $[u,p]\approx[15\pi,0]$. Each data point was obtained from $10^3$ trajectories, each simulated for $10^4$ driving periods. The chaotic Type-II trajectories are faster than those from the other chaotic region, so the velocity decreases with decreasing permeability of the cantorus.  }
\end{figure}

Before the driving amplitude is large enough to stabilize any of these new (quasi-) periodic Type-II trajectories, there is a long intermediary range of driving amplitudes during which the two chaotic regions are separated from each other by a permeable cantorus (i.e. an unstable KAM-torus). 
A corresponding phase space is shown in Fig. \ref{figure4}(d). 
The presence of the cantorus allows for an appealing dynamics of the chaotic trajectories. 
When they cross the cantorus, they switch between two different kinds of chaotic motion. 
An example for such a trajectory is shown in Fig. \ref{figure5}(a) for a driving amplitude of $E=500$. 
The plotted trajectory $u(t)+2\pi Mt$ has a negative (or positive) slope if the particle is moving faster (or slower) around the torus than the rotating driving field. 
The particle in the figure starts in the chaotic region outside of the cantorus barrier (i.e. it is not a Type-II trajectory). 
In this region it will (on average) be too slow to move in phase with the driving field. 
Once it crosses the cantorus, the dynamics become that of a chaotic Type-II trajectory. 
This is highlighted by the inset, which zooms into a small region of the trajectory during which the particle crosses the cantorus, briefly becomes a chaotic Type-II trajectory, and then crosses the cantorus again into the other chaotic region. 
The times particles spend as chaotic Type-II trajectories follow a power law with a critical exponent that depends on the driving amplitude and the permeability of the cantorus.

The permeability of the cantorus does not simply decrease with the driving amplitude until the two chaotic regions are separated from each other. 
It switches multiple times between being more or less permeable before the driving amplitude is large enough to separate the two chaotic regions. 
This is demonstrated in Figs. \ref{figure5}(b-d). 
They each show the PSOS of six trajectories (with the same initial conditions $[u,p]$ used for each figure) for various driving amplitudes. 
For $E=300$ and $E=400$, the two regions are almost separated from each other and in both cases only one of the trajectories manages to cross the cantorus. 
Despite the vast difference in driving amplitudes, there is very little difference in the permeability of the cantorus. 
In contrast, for $E=500$ all of the trajectories switch frequently between the two regions. 
In this case, the presence of a cantorus is not even obvious from the phase space alone. 
Only when looking at the individual trajectories (such as the one shown in Fig. \ref{figure5}(a)), we can distinguish between the different chaotic dynamics of the two chaotic regions. 

The average velocity is different for both chaotic regions, and we shall use this to analyze the split-up of the chaotic region. 
This is shown in Fig. \ref{figure5}(e). 
It shows the average velocity $v_{av}$ as a function of the driving amplitude. 
Each data point was obtained from $10^3$ trajectories started in the chaotic region around $[u,p]\approx[15\pi,0]$, and with simulation times of $10^4$ driving periods for each trajectory. 
Note, that for very low $E$, when the Type-III trajectories for $p>0$ and $p<0$ are separated by invariant spanning curves (i.e. Type-I trajectories), we chose initial conditions with $p<0$, leading to some bias in the data for very low $E$.
Note also, that the curve may slightly change for different simulation times, if the switch time for the cantorus crossing exceeds the simulation time.

At first, for very low $E$, $v_{av}$ decreases with increasing driving amplitude which is caused by a combination of Type-III trajectories disappearing with increasing $E$, and a `bias' in our initial conditions (compare $v_{av}$ in Fig. \ref{figure3}(e) and discussion thereof). 
Then, $v_{av}$ will (mostly) increase  with increasing driving amplitude until $E\approx270$. 
This increase is due to the chaotic sea expanding and changing its mean momentum.
Above $E\approx270$, the cantorus appears and splits the chaotic region in two, resulting in a sharp drop of $v_{av}$. 
From then on, there are peaks in $v_{av}$ whenever the trajectories can frequently switch between the two chaotic regions: (e.g. the plateau around $E=500$). 
Around $E\sim900$, the cantorus stabilizes into periodic Type-II trajectories and the two chaotic regions become fully separated from each other.

Similar to the phase space splitting for low driving amplitudes discussed in Sec. \ref{sec:driveX}, this split likely originates from the two different scales of oscillations in the driving potential. 
The cantorus orbits are mainly stabilized due to the large scale oscillation $\sim qER\cos(\omega t-u/M)$ of the running wave, with the smaller oscillations $\sim qEr\cos(u)\cos(\omega t-u/M)$ acting as a perturbation that (for a wide parameter range) prevents the Type-II trajectories from stabilizing and becoming periodic. 
Due to the position dependence of the smaller oscillations, the perturbation is always stronger for trajectories that are tightly bound i.e. closer to the fixed point, than for those with greater variations of $\dot{u}(t)-v_{av}$. 
This `perturbation' increases with increasing the helix radius $r$ and therefore a larger helix radius requires larger driving amplitudes for the chaotic region to split-up. 
Similar to the discussion of Sec. \ref{sec:driveX}, the winding number $M$ changes the ratio of $r/R$ and the velocity of the Type-II trajectories. 
This can influence the general parameter regimes in which the split-up of the chaotic phase space region is encountered, however, we did not observe any changes in the underlying physics when varying the winding number $M$.

\section{Summary and Conclusion}
\label{sec:end}

We have investigated the dynamics of a charged particle confined to a toroidal helix which is exposed to external driving forces originating from a time-dependent electric field. 
The main results consist in the phenomenological description and understanding of two different mechanisms for the split-up of the chaotic phase space region - both with their own interesting consequences for the dynamics. 
We showed that for low driving amplitudes the two different spatial scales of oscillating potential lead to a split-up of the chaotic region around $p=0$. 
This prevents chaotic trajectories to invert the direction of their motion and leads to a consistent average velocity of $|v_{av}|\approx2\pi$ for all diffusive trajectories. 
Especially notable is that this split allows for chaotic particle trajectories with non-zero average velocity, even in a case where the spatio-temporal symmetries that are usually associated with chaotic transport are not broken by the driving field. 
Our understanding of this split and the resulting dynamics is certainly also of interest in the context of Brownian motors.

Specifically for driving with a circularly polarized field in the xy-plane, we found another mechanism for the split-up of the chaotic sea - this time splitting off a chaotic region in which particles are trapped in a valley of the driving potential.
Trajectories confined to this separate region of the phase space move around the torus in phase with the driving field and will have a consistent average velocity of $v_{av}=2\pi M$. 
Before this region is completely separated from the remainder of the chaotic sea, there is a very large range of driving amplitudes for which the trajectories can switch between the two chaotic regions by crossing a permeable cantorus. 
The probability of crossing the cantous fluctuates heavily with the driving amplitude. 
The origin of this separation has been identified as a small perturbation of the driving potential, that is most influential around the extrema of the running wave and vanishes in between those extrema.

The presented spit-ups of the chaotic phase space region are not unique to setups with confining forces and mainly depend on the different scales of oscillations in the driving potential. 
A realization of similar physics in a driven lattice with spatially varying forces, or with ultracold atoms in an optical lattice seem feasible. 
Furthermore, recent experiments have demonstrated the possibility of confining neutral atoms to a helical path \cite{reitz2012a}, however, in such setups, the realization of our driving forces may be a challenge.

\begin{acknowledgments}
The authors thank Aritra K. Mukhopadhyay for helpful discussions.
\end{acknowledgments}

\bibliography{txtest}

\end{document}